\documentclass[twocolumn,pre,floats,floatfix,aps,amsmath,amssymb,showkeys]{revtex4}
\usepackage{graphicx}
\usepackage{epstopdf}
\usepackage{color}

\DeclareGraphicsExtensions{.pdf,.eps,.png,.jpg,.mps}

\begin{document}

\title{Resonant Localized Modes in Electrical Lattices with Second Neighbor Coupling}

\author{Xuan-Lin Chen $^{1,2}$, Saidou Abdoulkary $^{3}$, P. G. Kevrekidis $^{4}$, L. Q. English $^{1}$}

\affiliation{$^{1}$Department of Physics and Astronomy, 
Dickinson College, Carlisle, Pennsylvania, 17013, USA}

\affiliation{$^{2}$Physics Department, Harbin Institute of Technology, Harbin 150001, Heilongjiang Province, China}

\affiliation{$^{3}$D\'epartement des Sciences Fondamentales, de Droit et des Humanit\'es, IMIP University of Maroua, P.O. Box 46, Maroua, Cameroon}

\affiliation{$^{4}$Department of Mathematics and Statistics, University of Massachusetts, Amherst, Massachusetts 01003, USA}

\date{\today}

\begin{abstract}
  We demonstrate experimentally and corroborate
  numerically that an electrical lattice with nearest-neighbor and
  second-neighbor coupling can simultaneously support long-lived
  coherent structures in the form of both standard intrinsic localized modes (ILMs), as well as resonant ILMs. In the latter case,
  the wings of the ILM exhibit oscillations due to resonance with a degenerate plane-wave mode. This kind of localized mode has also been termed nanopteron. Here we show experimentally and using realistic simulations of the system that the nanopteron can be  stabilized via both direct and subharmonic driving. In the case of excitations at the zone center (i.e., at
  wavenumber $k=0$), we observed stable ILMs, as well as a periodic localization pattern in certain driving regimes. In the zone boundary case
  (of wavenumber $k=\pi/a$, where $a$ is the lattice spacing),
  the ILMs are always resonant with a plane-wave mode, but can
  nevertheless be stabilized
  by direct (staggered) and subharmonic driving. 
\end{abstract}


\maketitle

\section{Introduction}
Intrinsic localized modes (ILMs), also known as discrete breathers (DB), have been studied in a great variety of physical systems whose only requirements are that they (i) be spatially discrete and (typically) periodic, and (ii) exhibit nonlinearity \cite{campbell04,gorbach}. While some of the prototypical
early studies arose in the context of vibrational-energy localization in atomic lattices~\cite{st,page}, relevant considerations quickly expanded
--involving also experimental work-- to spin chains \cite{Lai1999}, and later to fabricated photonic lattices, Bose-Einstein condensates, Josephson-junction arrays, and MEMS systems, to name just a few examples~\cite{fleischer03,
  smerzi, trias, sato06}. More recently,
materials systems such as granular chains~\cite{db10,db14,vakakis}, as well
as
electrical lattices~\cite{sato07, lars08, Shige2018-electronic}
have attracted some interest as media supporting self-localized modes.

One thing that was realized early on was that ILMs could only exist when their frequencies fell outside of the spectrum of linear, extended-wave modes. In fact, even the overtones of the ILM were forbidden to overlap any plane-wave mode frequency. This was established rigorously for Hamiltonian lattices \cite{macaub,flach94}. 
On the other hand, it was also demonstrated that discrete breather-like modes could exist in certain circumstances even when they did intersect part of the linear dispersion curve. In these cases, the mode energy was the highest in a very narrow central region of space, but it did not strictly go to zero with distance from that center. Instead, the tails of the ILM were found to exhibit small spatial oscillations consistent with the wavenumber of the degenerate plane-wave mode. This type of phenomenon was reported, for instance, in numerical studies of spin-wave localization in spin chains with second-neighbor Heisenberg interactions \cite{Lai97a,Lai97b,Kiselev98,Huang1998-ferromagnetic}. The authors named this type of self-localization an ``intrinsic localized spin-wave resonance.'' No corresponding experimental observations were obtained,
to the best of our knowledge, in this setting.
The term {\it nanopteron} has also been used to signify a soliton-like solution featuring a small oscillatory background, and additionally discrete versions of such a solitary wave have been discussed \cite{savin,iooss};
such structures are also referred to as ``weakly nonlocal solitary waves''
and a summary of (early) theoretical and numerical efforts along this direction
was given in Ref.~\cite{boyd}. However, experimental studies of such excitations remain very limited; the only systematic example that we are
aware of concerns the propagation of weakly nonlocal traveling waves
in a woodpile lattice in the work of Ref.~\cite{kim}. Coherent structures
in the form of weakly localized discrete breathers will be a focal point
of the present study.

Another key feature of the present work is the consideration of
beyond nearest-neighbor interactions, through the inclusion of second
neighbors. 
The importance of incorporating such interactions, when studying
energy localization has also been reported in a variety of other discrete extended systems, many of which can be modeled by the nonlinear Schr\"odinger equation \cite{kevrekidis2003-DNLS, tang2017-k2k4, yao2015-betaFPU}. A particular
proposal of this type, through the consideration of so-called zigzag
lattices in nonlinear optics~\cite{efremidis} subsequently
found remarkable experimental realizations in the context of
femtosecond laser-written waveguide arrays~\cite{szameit}. Subsequent
studies in that context went beyond the study of localization and
also towards the observation of (anharmonic) Bloch oscillations by
employing zigzag waveguide arrays in which the second-neighbor coupling
could be controllably tuned~\cite{szameit2}. The study of the relevant
phenomenology remains an active topic in recent
investigations~\cite{pgk,tiziano}; it will also be central to
the discrete breather phenomenology in the electrical lattices
presented herein.

In this paper, we report on experimental and numerical observations of ILMs in an electrical lattice with nearest neighbor and second-neighbor coupling.
A key distinguishing feature of our work with
respect to earlier optical realizations
is that contrary to the scenarios of the latter type~\cite{szameit,szameit2}
where the second-neighbor coupling is predominantly linear,
in our case, it is fundamentally
nonlinear, as will be evident in the mathematical model below.
Moreover, 
the theoretical model contains a nonlinear coupling of each node to its
nearest neighbors
comparable to the onsite nonlinearity, a feature also rather uncommon
in the optical setting; see, e.g., the relevant discussion of~\cite{johansson}.
It is shown that this lattice can support two types of
long-lived, experimentally observable
localized modes: the standard ILM (whose wings asymptotically
approach zero), and the resonant ILM or nanopteron (whose wings manifest
oscillations). The former branches off from the zone-center (ZC) mode
(k=0), whereas the latter derives from the zone-boundary (ZB) mode (k=$\pi/a$). 
For a first analytical exploration, we deploy a simplified model of the lattice that uses harmonic and square-anharmonic onsite terms, approximating the full nonlinearity of the electrical unit-cell LC-resonator fairly well, while ignoring the resistive currents through the diode. In this context, we can numerically
approximate the solutions, and show that both standard ILMs as well as resonant ILMs each come in two varieties of even- and odd symmetries. 

On the experimental side, both types of localized modes are generated and stabilized with an external driver via modulational instability. In the case of the standard ILM (here also referred to as ZC ILM), the driving signal is spatially homogeneous with a frequency just below the bottom of the linear dispersion curve. For resonant ILMs or nanopterons (also referred to as ZB ILMs), we employ a spatially staggered driver just below the ZB mode in frequency, as well as subharmonic driving. Realistic dynamical  simulations are performed, and these too indicate that such excitations can be generated via modulational instability of the relevant extended mode.

\section{The System and Theoretical Model}

\begin{figure}[tbh]
\centering
\includegraphics[width=3.5in]{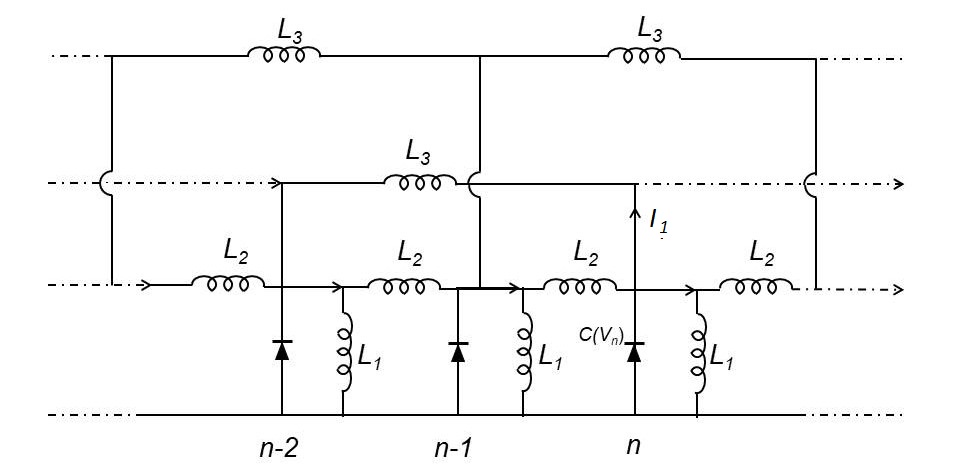}\textbf{}
\caption{Schematic representation of the coupled nonlinear transmission line.} \label{setup}
\end{figure}
The nonlinear lattice shown in Fig. \ref{setup} consists of N unit cells
each comprised of a linear inductor $L_{1}$ in parallel with a nonlinear capacitor $C(V)$ and coupled via inductances $L_{2}$ and $L_{3}$. The nonlinear capacitor is typically realized by a (reverse-biased) varactor diode with a capacitance given as a function of the differential voltage $V_{n}$ across the capacitor. 

For reasons of analytical tractability, we begin the discussion by exploring a simplified dynamical model, where the nonlinear element is considered to be strictly capacitive which, in contrast to a diode, precludes the possibility of resistive currents flowing through the element. This is the primary limitation of this model in modeling the real varactor diode used in the experiment. Later, we will come back to a more realistic, yet also more complex, model that we use as the basis for direct simulations. 

Proceeding with the simpler model, and using the Kirchhoff node rule, the equations for the unit-cell circuit become,
\begin{eqnarray}
\label {eq.2}
I_{n}=I_{n+1}+I_{1}+\frac{d Q_{n}}{dt}+I_{L1} \\ \nonumber
L_{3}\frac{d I_{1}}{dt}=2 V_{n}-V_{n-2}-V_{n+2},
\end{eqnarray}
where $I_{L_{1}}$ is the current through the inductor $L_{1}$, and $Q_{n}$ is the charge on the $n^{th}$ capacitor.  Combining Eq.~(\ref{eq.2}) with the auxiliary equations $L_{1}dI_{L_{1}}/dt=V_{n}$, $L_{2}dI_{n}/dt=V_{n-1}-V_{n}$, we obtain the following governing equation,
\begin{eqnarray}
\label {eq.3}
\frac{d^{2}Q_{n}}{dt^{2}}&=&-\frac{V_{n}}{L_{1}}+\frac{1}{L_{2}}(V_{n+1}+V_{n-1}-2V_{n})\\ \nonumber
&+&\frac{1}{L_{3}}(V_{n+2}+V_{n-2}-2V_{n})
\end{eqnarray}
In order to get the relationship between $Q_{n}$ and $V_{n}$, a particular functional form of $C(V)$ has to be specified, which can then be integrated with respect to the voltage to yield $Q(V)$. For instance, if an exponential dependence is assumed, $C(V_{n})=C_{0}\exp(-\alpha V_{n})$, then integration and subsequent inversion of the resulting expression will produce,
\begin{eqnarray}
\label{eq.4}
V_{n}=-\frac{Q_{0}}{C_{0}}\ln(1-\frac{Q_{n}}{Q_{0}}),
\end{eqnarray}
where $Q_{0}=C_{0}/\alpha$. 
However, this relationship when applied to a single oscillator does not generate a soft nonlinearity. It is an experimental fact about the varactor diodes that they are characterized by soft nonlinearity. Even truncating the Taylor expansion of Eq.~(\ref{eq.4}) after the cubic term would still yields an oscillator whose frequency increases with amplitude. To achieve soft nonlinearity we focus
on the scenario where the square term dominates over the cubic (see also Ref.~\cite{sato07}). It turns out that keeping only the first two terms provides a frequency-amplitude dependence that is fairly close to the experimental data. This approximation was also used in Ref.~\cite{lars2}. Thus, we assume,
\begin{eqnarray}
\label{eq.5}
V_{n}=\frac{Q_{0}}{C_{0}}(q_{n}+\frac{1}{2}q_{n}^{2}),
\end{eqnarray}
with $q_{n}=Q_{n}/Q_{0}$, Eq.(\ref{eq.4}).

Now inserting Eq.~(\ref{eq.5}) into Eq.~(\ref{eq.3}), we obtain
\begin{flalign}
\label {eq.6}
&\frac{d^{2}q_{n}}{d\tau^{2}}=-(q_{n}+\frac{1}{2}q_{n}^{2})\\ \nonumber
&+\gamma[(q_{n+1}+\frac{1}{2}q_{n+1}^{2})+(q_{n-1}+\frac{1}{2}q_{n-1}^{2})-2(q_{n}+\frac{1}{2}q_{n}^{2})]\\ \nonumber
&+\delta[(q_{n+2}+\frac{1}{2}q_{n+2}^{2})+(q_{n-2}+\frac{1}{2}q_{n-2}^{2})-2(q_{n}+\frac{1}{2}q_{n}^{2})]
\end{flalign}
where $\tau=\omega_{0}t$ is effective time parameter, $\omega_{0}^{2}=1/(L_{1}C_{0})$ is the frequency of uniform mode, and $\gamma=L_{1}/L_{2}$ and $\delta=L_{1}/L_{3}$ are inductor ratio parameters.

Anticipating the possibility of dynamic anharmonic modes, we set a trial solution \cite{Bickham1993-cubic} as
\begin{eqnarray}
\label{eq.7}
q_{n}(\tau)=A(\phi_{n}(\tau)\cos(kna+\tilde{\omega} \tau)+\xi_n(\tau))
\end{eqnarray}
where $A$ is the maximum amplitude, $a$ is the lattice spacing, $\cos(kna+\tilde{\omega} \tau)$ is a moving carrier wave with $\tilde{\omega}=\omega/\omega_{0}$, while $k$ and $\omega$ are its wave vector and frequency, respectively. $\phi_{n}(\tau)$ and $\xi_n(\tau)$ are the vibrational envelope and the
non-oscillatory
displacement, respectively. For initial displacements, $\tau=0$, $\phi_{n}\cos(kna)$ is often known as ac displacement, and $\xi_{n}$ as dc displacement.
We set $A=1$ for simplicity, then insert Eq.~(\ref{eq.7}) into Eq.~(\ref{eq.6}), and using the rotating wave approximation method (RWA),
while also assuming that $\phi_n$ and $\xi_n$ are $\tau$-independent,
we can obtain the cosine terms
\begin{eqnarray}
\label{eq.8}
&\phi_{n}[-1+\tilde{\omega}^{2}-2\delta-2\gamma-(1+2\delta+2\gamma)\xi_{n}]\\ \nonumber
&+\gamma[\phi_{n-1}(1+\xi_{n-1})+\phi_{n+1}(1+\xi_{n+1})]\cos(ak)\\ \nonumber
&+\delta[\phi_{n-2}(1+\xi_{n-2})+\phi_{n+2}(1+\xi_{n+2})]\cos(2ak)=0,
\end{eqnarray}
and the static displacement terms
\begin{eqnarray}
\label{eq.9}
\delta \phi_{n-2}^{2}+\gamma \phi_{n-1}^{2}-\phi_{n}^{2}-2(\delta+\gamma)\phi_{n}^{2}+\gamma \phi_{n+1}^{2}\\ \nonumber
+4\gamma \xi_{n-1}+2\gamma \xi_{n-1}^{2}+\delta[\phi_{n+2}^{2}+2\xi_{n-2}(2+\xi_{n-2})-8\xi_{n}]\\ \nonumber
-4\xi_{n}-8\gamma \xi_{n}-2\xi_{n}^{2}-4\delta \xi_{n}^{2}-4\gamma \xi_{n}^{2}+4\gamma \xi_{n+1}\\ \nonumber
+2\gamma \xi_{n+1}^{2}+4\delta \xi_{n+2}+2\delta \xi_{n+2}^{2}=0.
\end{eqnarray}
Equations (\ref{eq.8}) and (\ref{eq.9}) are used in the next section to find numerically {\it approximate} breather solutions that can be tested against the
experimental and numerical results. 

\begin{figure}
\centering
\includegraphics[width=3in]{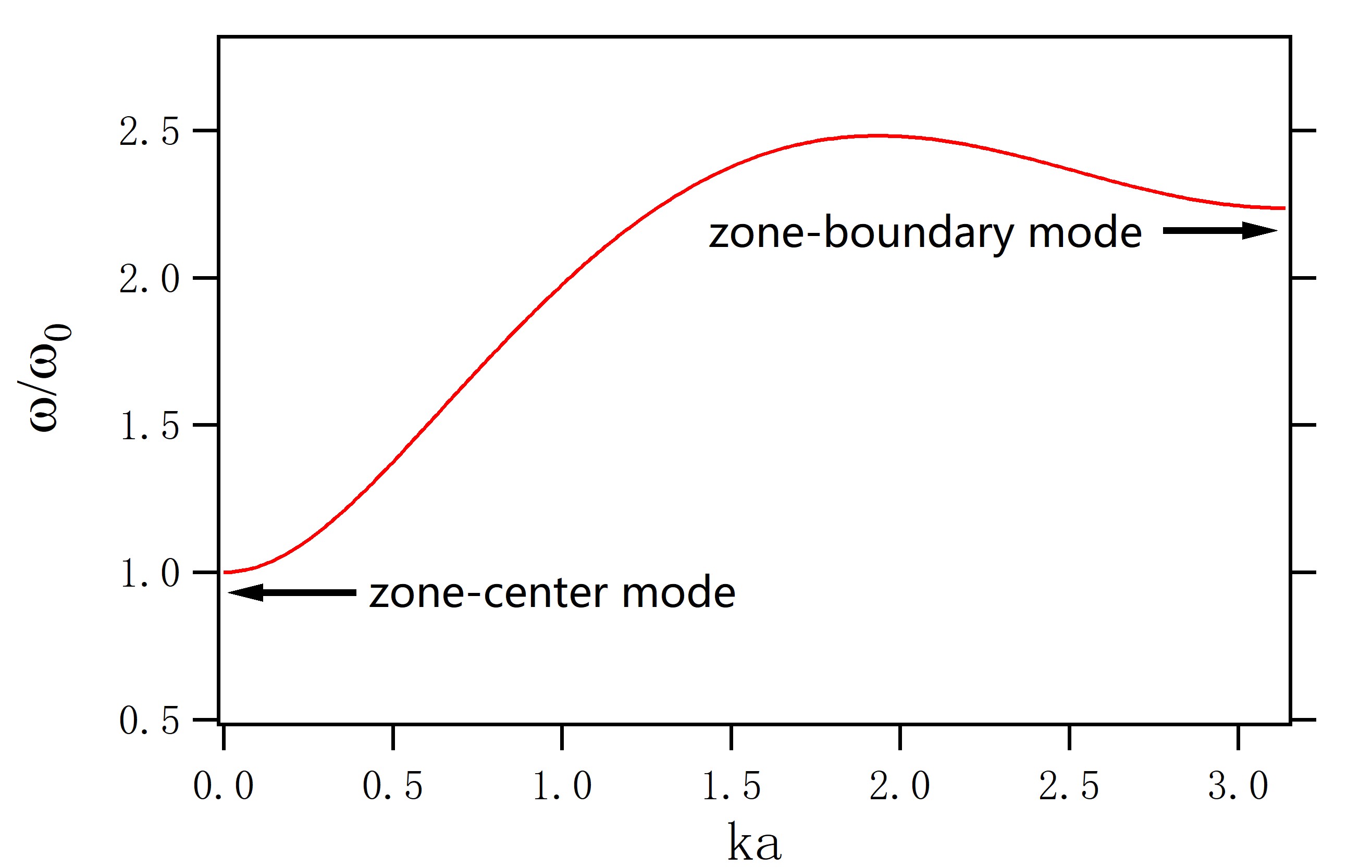}\textbf{}
\caption{Dispersion relation showing frequency as a function of $ka$. The standard ILMs and resonant ILMs should be found at $ka=0$ and $ka=\pi$ under the dispersion curve, respectively.} \label{Fig.4}
\end{figure}

The linear limit of Eq.~(\ref{eq.6}) yields the following linear dispersion relation:
\begin{equation}
\tilde{\omega}^2(k)=1+4\gamma \sin^2 \left(\frac{ka}{2}\right)+4\delta \sin^2(ka)
\end{equation}
The dispersion curve is shown in Fig.~\ref{Fig.4}. A frequency gap exists below the zone-center (uniform) mode due to the inductor to ground in each unit cell, but additionally a quasi-gap also appears at the zone-boundary due to the second-neighbor inductors.
By this term, we mean the frequencies below those of the ZB at $k=\pi/a$, where no nearby modes exist in k-space.
This creates the possibility that resonant ILMs could be found arising from the ZB mode. Note that the curvatures of the band at the two zone edges are both positive (for $\delta>0$).

The analysis above was based on a simplified model of the varactor diode as a nonlinear capacitor. A fuller description of the diode must also account for current going through it, thereby including resistive dissipation. Reference \cite{faustino} derives equations that include this feature as well as external driving, which when modified to our system with second neighbor coupling read,
\begin{eqnarray}
\label{eqline}
c(v_n) \frac{d v_n}{d \tau} & = & y_n-i_D(v_n)+ \frac{\cos(\Omega \tau)}{RC_0\omega_0} -
\nonumber  \\ & &
\left(\frac{1}{R_l}+\frac{1}{R}\right) \frac{v_n}{\omega_0 C_0} \nonumber, \\[2.0ex]
 \frac{d y_n}{d\tau} & = & -v_n + \frac{L_1}{L_2}\left(v_{n+1}+v_{n-1}-2v_n\right) \\ \nonumber &+&\frac{L_1}{L_3}\left(v_{n+2}+v_{n-2}-2v_n\right) ,  \\ \nonumber
\end{eqnarray}
where $R$ is the driving resistor of 10 k$\Omega$, and $R_l$, $c(v)$, $i_D(v)$ are given in Ref.\cite{faustino} as,
\begin{equation}
 i_D(V)= -\frac{I_s}{\omega_0 C_0 V_d} \exp(-\beta V),
\nonumber
\end{equation} 
with $\beta=38.8$~V$^{-1}$ and $I_s=1.25 \times 10^{-14}$ A. Furthermore, $v=\frac{V}{V_d}$, $c=\frac{C}{C_0}$, and
\begin{equation}
 C(V)=
\begin{cases} \begin{matrix}
C_{v}+C_{w}(V')+  C(V')^2  & \mbox{if} \quad V \leq V_{c}, \\[2.0ex]
  C_0 e^{-\alpha V} & \mbox{if} \quad V > V_{c}.
\end{matrix} \end{cases}
\nonumber
\end{equation}
Here, $V'=(V-V_c)$, $\alpha=0.456$~V$^{-1}$,
$C_{v}=C_0\exp(-\alpha V_{c})$,  $C_{w}=-\alpha C_{v}$, $C=100$~nF/V$^2$, and
$V_{c}=-0.28$~V.

Equation (\ref{eqline}) represents the
more accurate variant of our model that we will use to
compare more directly to experimental results. Importantly, it encompasses
the driving and damping aspects inherent in the electrical lattice. Hence,
we utilize in what follows the simplified model of Eq.~(\ref{eq.6})
to obtain insight on the potential existence and structural form of
the breathers; then, we explore whether such structures survive
in the more realistic setting containing the driving/damping.

\section{Results and Discussion}
\subsection{Approximate (RWA-based) numerical solutions in the undriven, undamped lattice}
\begin{figure}[tbh]
\centering
\includegraphics[width=3.5in]{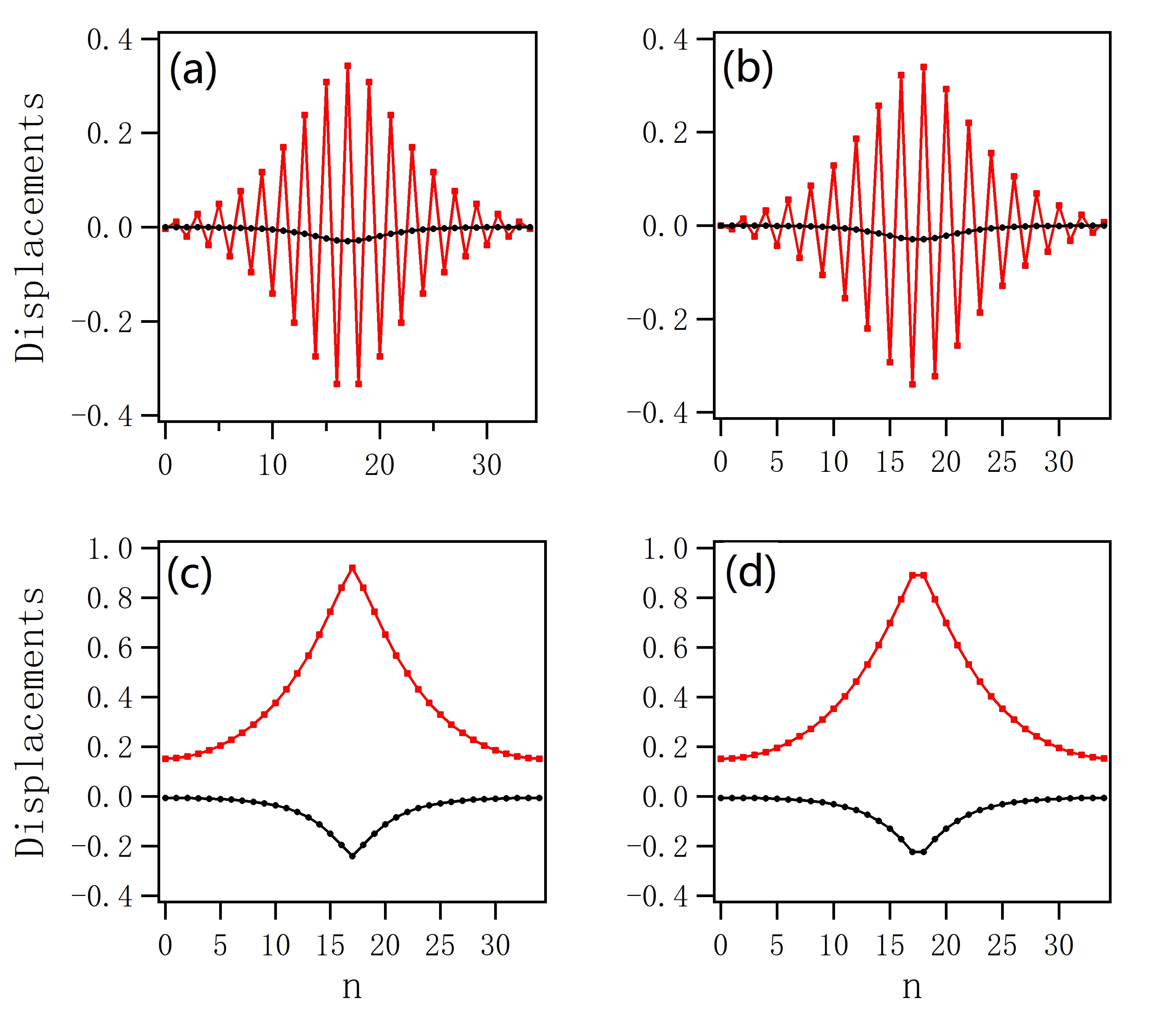}\textbf{}
\caption{Comparison of ac and dc displacements for {\it approximate}
  ILMs, as obtained within the RWA
  from the solution of Eqs.~(\ref{eq.8}) and (\ref{eq.9}). (a) odd ZB (resonant) ILM ($\tilde{\omega}=2.22$), (b) even ZB ILM ($\tilde{\omega}=2.22$), (c) odd ZC ILM ($\tilde{\omega}=0.95$), and (d) even ZC ILM ($\tilde{\omega}=0.95$). The filled squares (red) are for ac displacements, and the filled circles (black) show the dc displacements.} \label{Fig.1}
\end{figure}

We now employ the Newton-Raphson method to determine the solutions of
Eq.~(\ref{eq.8}) and Eq.~(\ref{eq.9}). If an initial guess of displacements is given, then solutions to these coupled, nonlinear algebraic equations
will be determined though an iterative process. Here, motivated by the experiments
below, we use the values $L_{1}=L_{2}=470\mu F$, $L_{3}=680\mu F$, so $\gamma=1$, $\delta\approx 0.7$. Both ZB ($ka=\pi$) and ZC ($ka=0$) ILMs are found in chains with 35 sites when imposing periodic boundary conditions, as shown in Fig.~\ref{Fig.1}. Notice that the dc-value is close to zero except near the
ILM centers.  

\begin{figure}[tbh]
\centering
\includegraphics[width=3.6in]{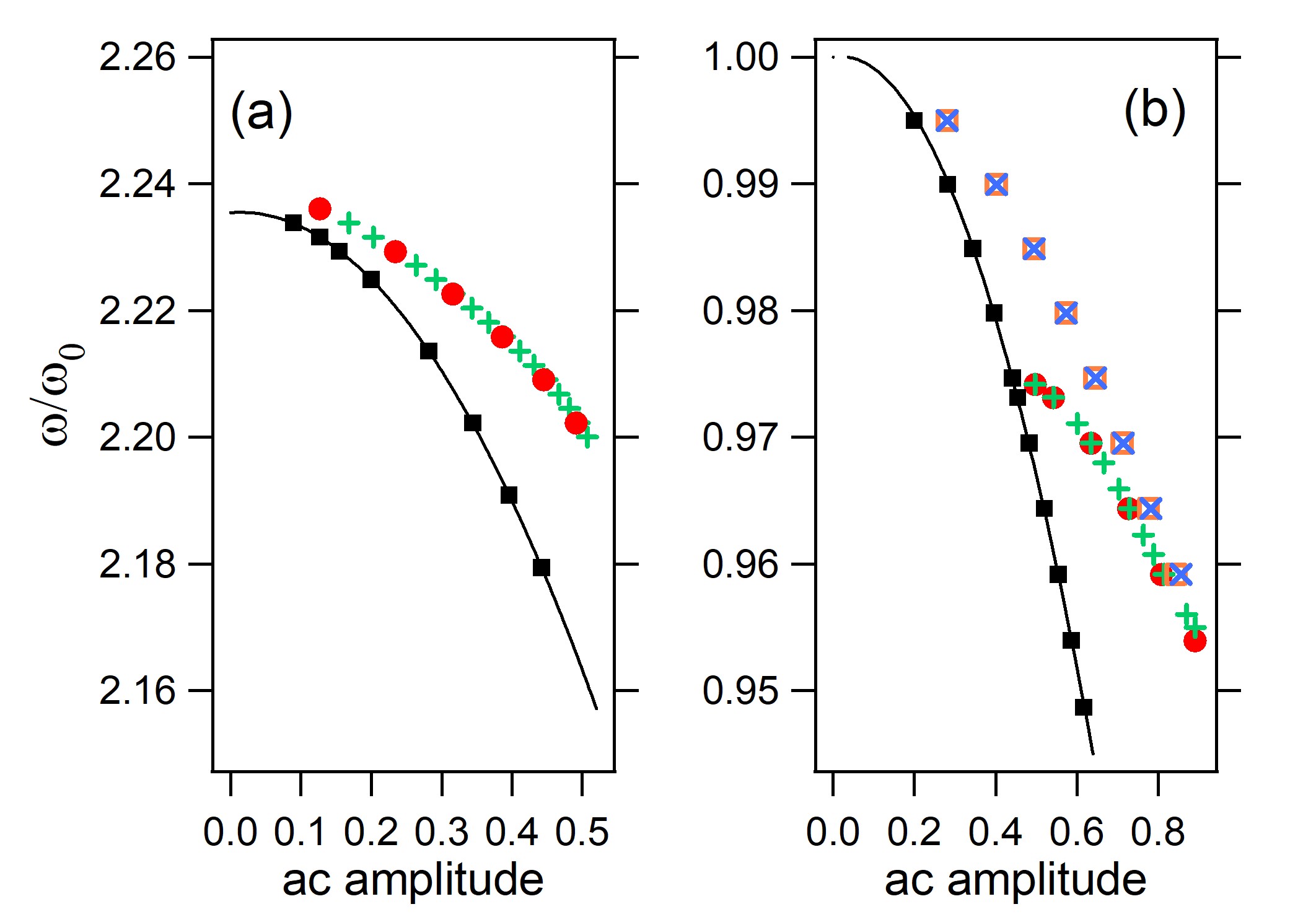}\textbf{}
\caption{The frequency of ILMs and associated plane-wave modes as a function of ac amplitude. (a) zone-boundary ILMs, and (b) zone-center ILMs. The filled squares (black) are for ka=0 and ka=$\pi$ plane-wave modes, respectively, which also have parabolic fitting lines (black). For N=35, the filled circles (red) are for even modes, and the plus-markers (green) show the odd modes. A second lattice-size, N=128 (squares and crosses), is also shown; here the ILMs are seen to branch off sooner.} \label{Fig.3}
\end{figure}

The relationship between the frequency of ILMs and their ac-amplitude is shown in Fig.~\ref{Fig.3}, which was obtained via continuation of the solution in $\omega$. The frequencies of odd ILMs are always a little higher than those of even ILMs with the same ac amplitude. What is interesting is that for smaller lattices, here N=35, the ILM branches off from the k=0 plane-wave (uniform) mode at a non-zero amplitude. For larger lattice sizes, this does not occur, as is shown in the figure for N=128. The qualitative reason is that for small lattices the zone-center ILM is wide enough that its wings do not reach a small amplitude within the extent of the lattice.

\begin{figure}[tbh]
\centering
\includegraphics[width=3.75in]{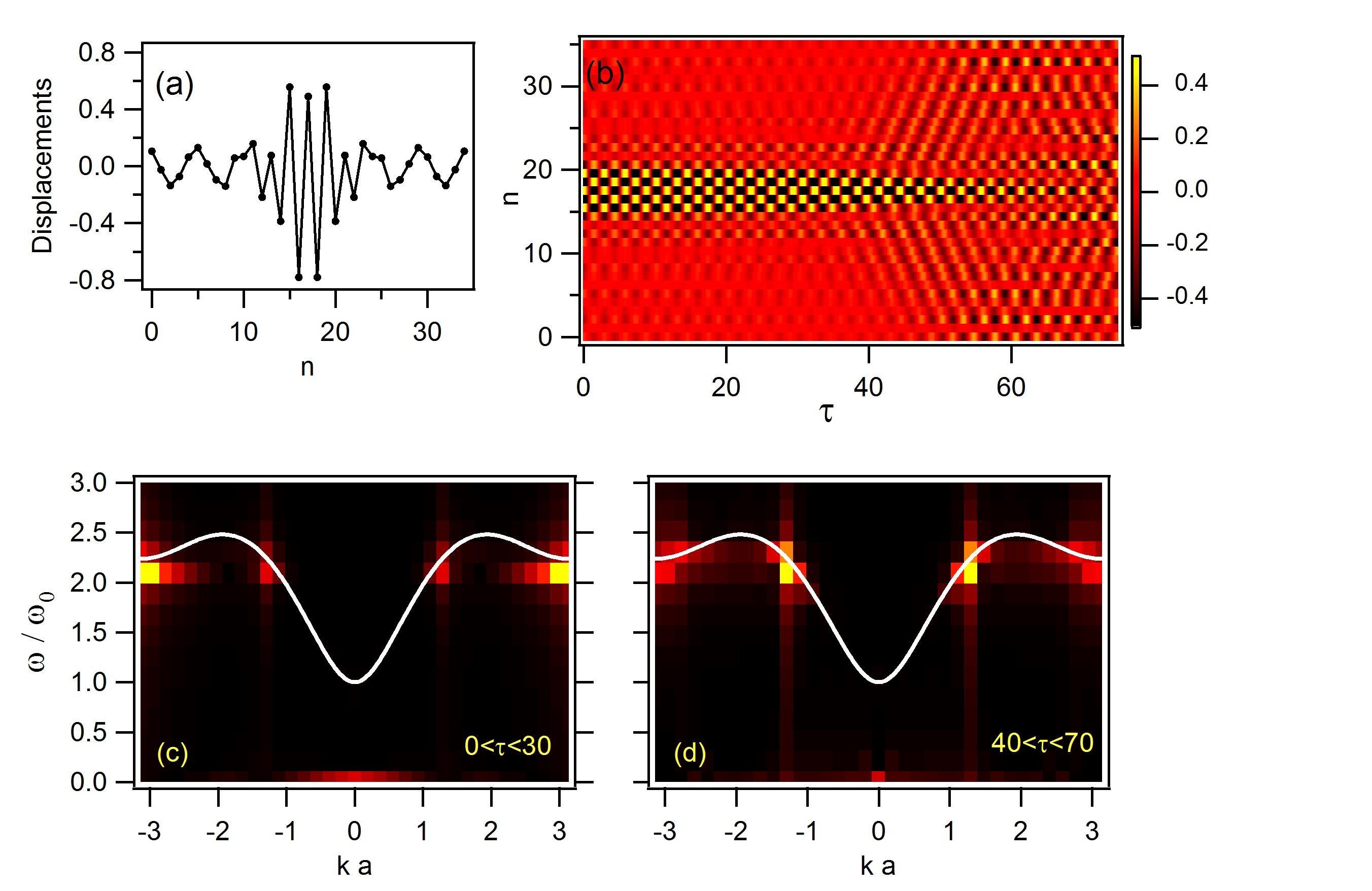}
\caption{(a) Displacements of an odd resonant ILM ($\tilde{\omega}=2.14$); (b)Time evolution of the odd resonant ILM at $\tilde{\omega}=2.14$; (c) Fourier transformation of the time evolution during $0<\tau<30$ and (d) during $40<\tau<70$, compared with the dispersion curve (white line).} \label{Fig.5}
\end{figure}

When the frequency of the odd ZB ILM becomes lower, and its amplitude higher, a strong resonance appears between our approximate breather profiles,
as obtained by Eqs.~(\ref{eq.8}) and (\ref{eq.9}) and shown in Fig.~\ref{Fig.5}(a), and the degenerate linear plane-wave mode of the dispersion curve. Figure \ref{Fig.5}(b) shows the numerical evolution using the RWA output as an initial condition. This waveform does not appear to be
robust when simulated using the full governing equations. The time-evolution
of Eq.~(\ref{eq.6}) using a Runge-Kutta algorithm is shown in Fig.~\ref{Fig.5}(b). We see that the initial waveform propagates for a number
of periods, but eventually disintegrates by radiating energy into the degenerate extended modes. This is also clearly seen in the two-dimensional Fourier transform graphs: Fig.~\ref{Fig.5}(c) corresponds to the early times (from 0 to 30 on the $\tau$ axis, before full disintegration), Fig.~\ref{Fig.5}(d) shows the later time window (between 40 and 70 on the $\tau$ axis). Comparing these two images, we see clearly that energy is transferred from the ZB ILM to the resonant extended mode. In this way the approximate, as well as resonant nature of the
relevant ZB waveform is manifested. For smaller amplitude ILMs, this transfer also happens, but at a slower rate, so that they appear robust for longer periods of time. At the ZC, while the ILM has no direct resonance, its overtones still intersect the linear spectrum. The RWA cannot account for this process, but direct simulations also indicate a slow decay of the long-lived localized mode.


\subsection{Experimental Results and Numerical Simulations}
We have constructed the lattice shown in Fig.~\ref{setup} of 32 nodes with $L_1=470 \mu H$, $L_2=680 \mu H$, $C_0=800 pF$. Care was taken to select inductors with very similar inductances to reduce spatial inhomogeneity. These component values yield a frequency of $f_0=\omega_0/(2\pi) \cong 260 kHz$. The lattice takes the shape of a ring to eliminate boundaries. This also practically
implies the generic implementation
of periodic boundary conditions in the context
of numerical computations.

\begin{figure}[tbh]
\centering
\includegraphics[width=3.0in]{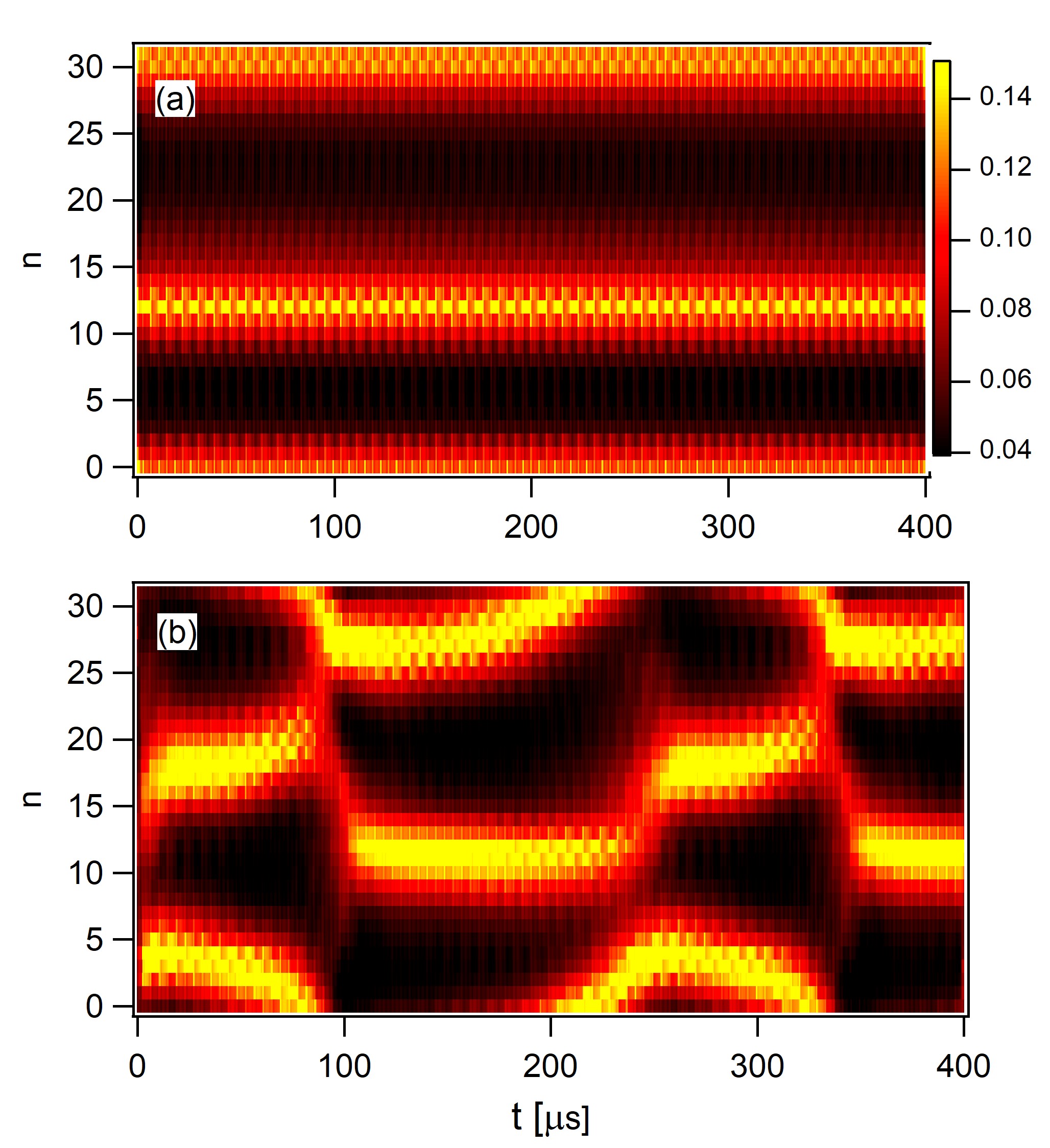}
\caption{Experimental data illustrating the averaged square voltage
  evolution (color-bar on the right) at each node as a function of time. (a) The spatially uniform driver was set to 249 kHz in frequency at an amplitude of 2.0V. Two broad zone-center ILMs are generated. (b) The driver frequency is decreased to 239 kHz. The system develops a repeating pattern of two
  ILMs that interact multiple times.}
\label{exp1}
\end{figure}
In order to access the area of interest in k-space, we drive the system at all lattice nodes via 10 k$\Omega$ resistors connected to the top end of the diodes. The idea then is to directly excite plane-wave modes by having the driver match their frequency; alternatively, the driver can also be set to twice the mode frequency for subharmonic excitation - a method we employ later. For direct driving, we also have added some control of the driver wavenumber. In particular, in order to stimulate the ZB plane-wave mode, we tune the signal generator to about 580 kHz and introduce a phase shift of $\pi$ between neighboring nodes. This kind of spatially staggered driving can be easily accomplished by sending the original sinusoidal driving signal through an inverter (such as an inverting amplifier of gain 1) for half of the nodes, while using the direct signal for the other half.

Performing a frequency sweep with a spatially  homogeneous driving, only the ZC uniform mode is observed, whereas for spatially staggered driving,
as is natural to expect, we only couple to the ZB mode. Having demonstrated that the ZC and ZB extended modes can be excited in this fashion, we explore whether ILMs can also be generated and stabilized by these two types of driving.
In particular, we examine experimentally whether a temporally periodic driver
can couple to and sustain these modes. 

Figure \ref{exp1} shows that this indeed the case for the zone center when driving uniformly at each node in the experiment. The averaged square voltage density plot (see the color-bar) indicates that two fairly broad ILMs are generated - their width is enhanced by the presence of next-neighbor couplings.
The average square of the voltage is shown at each node, which was calculated by squaring all voltages and then using a box-car integration method with a time-window of one period.
The driver frequency here is set to 249 kHz, or $\omega_d=0.97 \omega_0$, and the amplitude is 2.0 V. When the driver frequency is decreased to 239 kHz, or $\omega_d=0.93 \omega_0$, a repeating pattern of two ILMs merging and splitting is observed. 

\begin{figure}[tbh]
\centering
\includegraphics[width=3.0in]{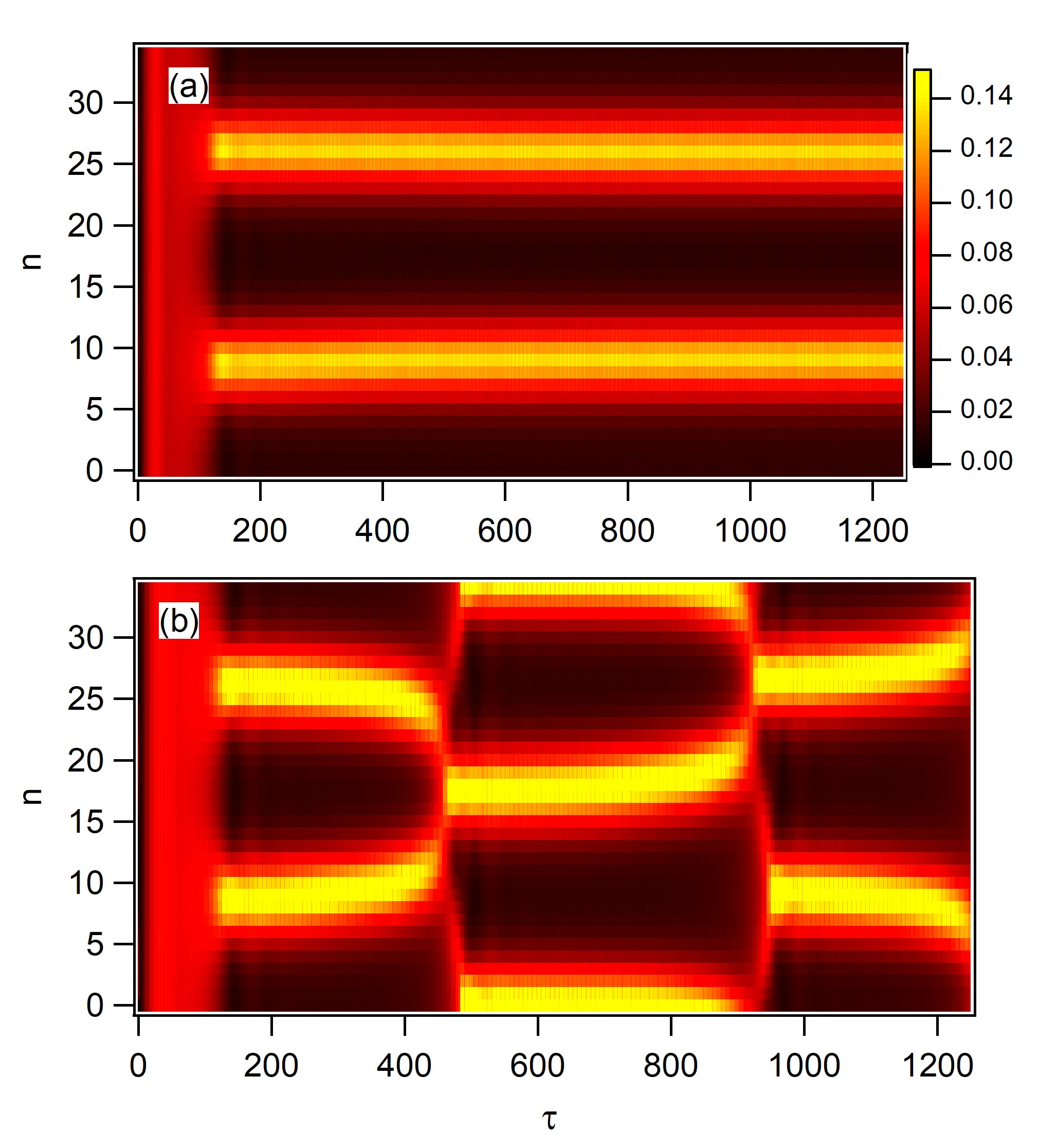}
\caption{The time evolution numerical results of averaged square voltage of ZC ILMs at an amplitude of 2.0V with a frequency of (a) $\tilde{\omega}=0.97$ and of (b) $\tilde{\omega}=0.93$.} \label{num1}
\end{figure}
Direct numerical simulations of Eq.~(\ref{eqline}) corroborate these experimental findings. Here, a  lattice of 35 nodes is initiated with small spatial noise; the time evolution of its averaged square voltage in the presence of driving is shown in Fig.~\ref{num1}. In Fig.~\ref{num1}(a), we can see two ZC ILMs and they seem quite robust at this driving frequency. When the frequency gets lower as shown in Fig.~\ref{num1}(b), the dynamics exhibits a periodic localization pattern, but otherwise persists. Thus, the simulations appear to faithfully reproduce the experimental observation of Fig.~\ref{exp1} at the identical driving parameters. The qualitative reason for the appearance of the dynamic localization pattern at these lower frequencies is that it represents a transition region to one ILM at even lower frequencies.

\begin{figure}[tbh]
\centering
\includegraphics[width=2.9in]{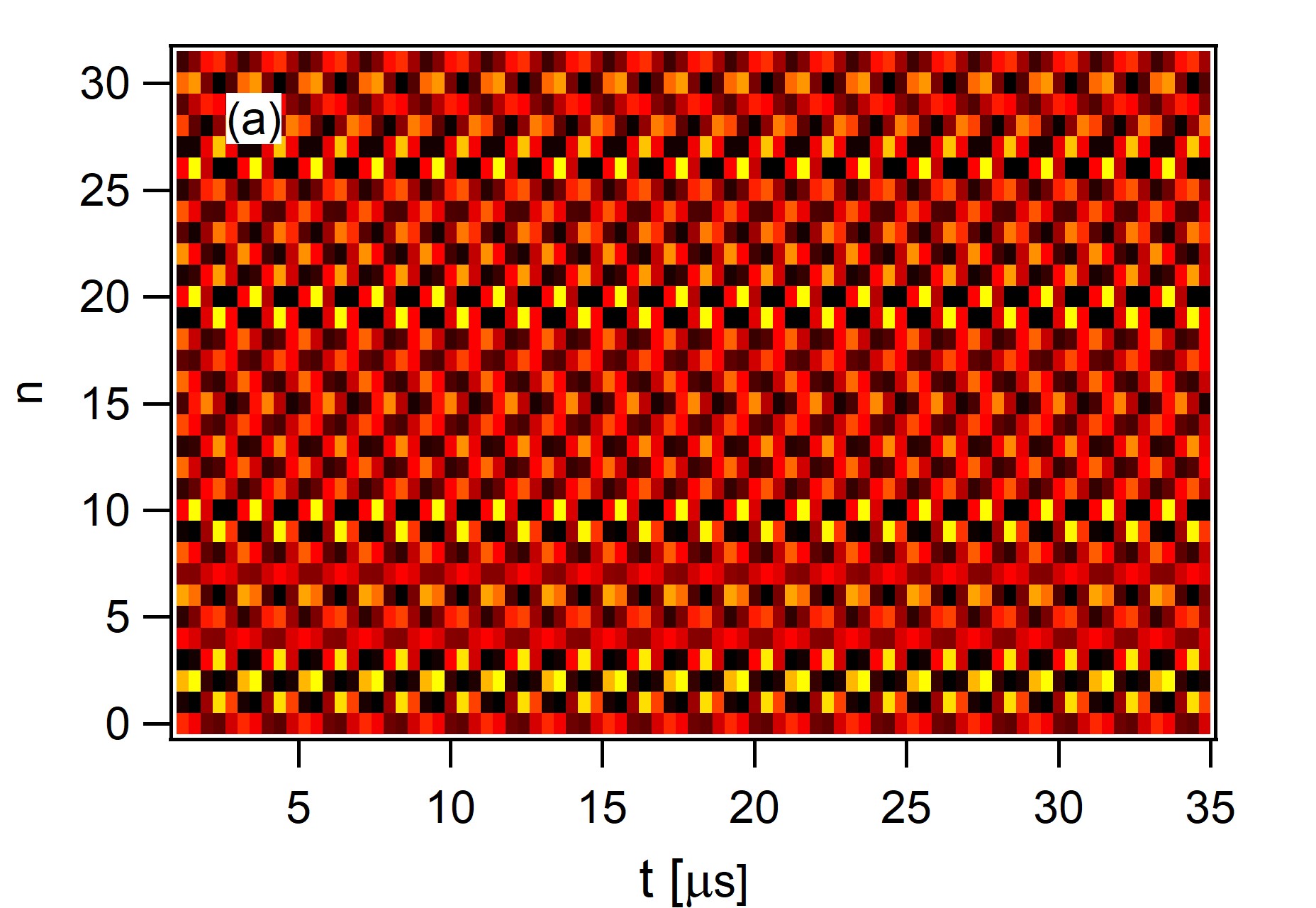}
\includegraphics[width=2.9in]{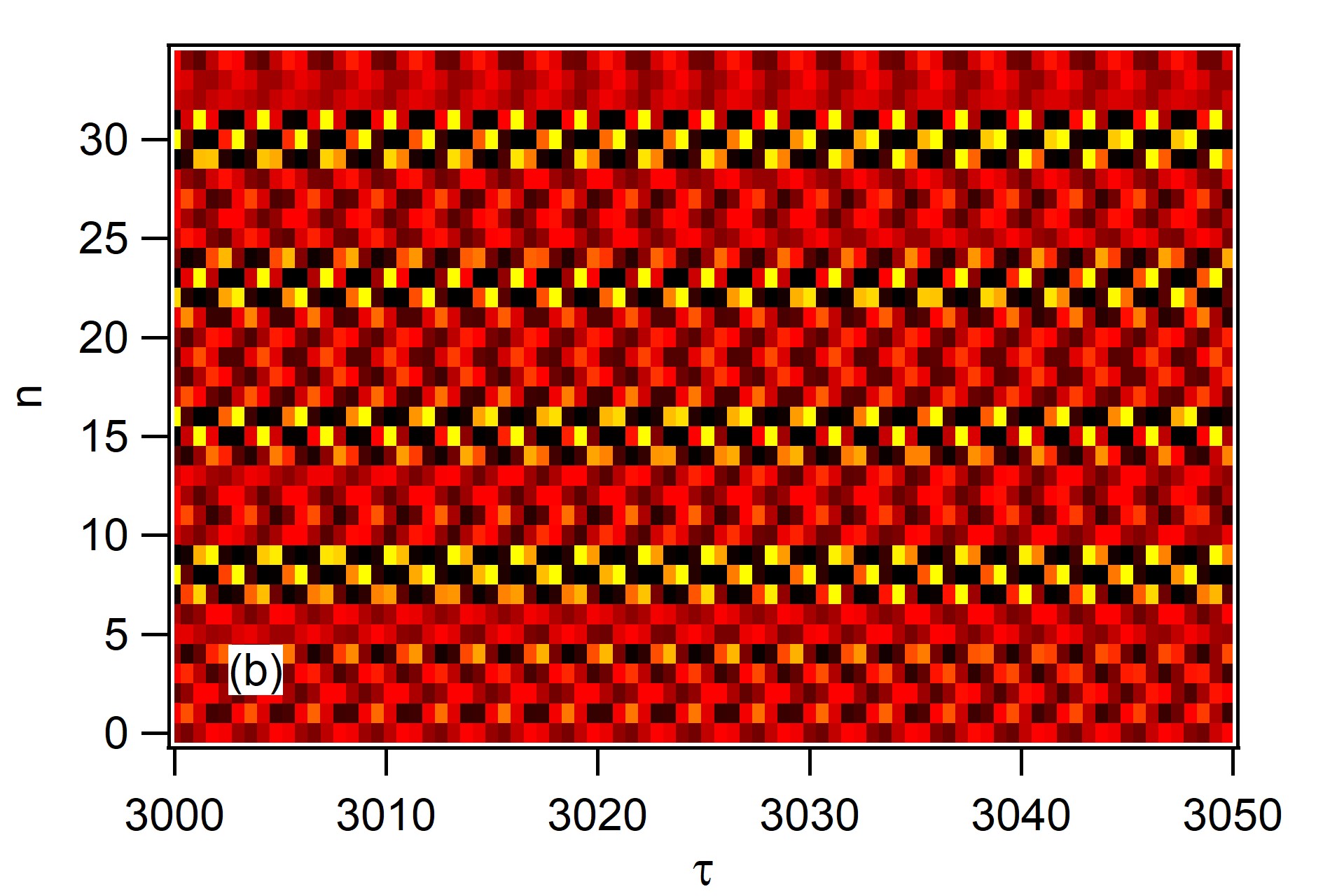}
\includegraphics[width=2.7in]{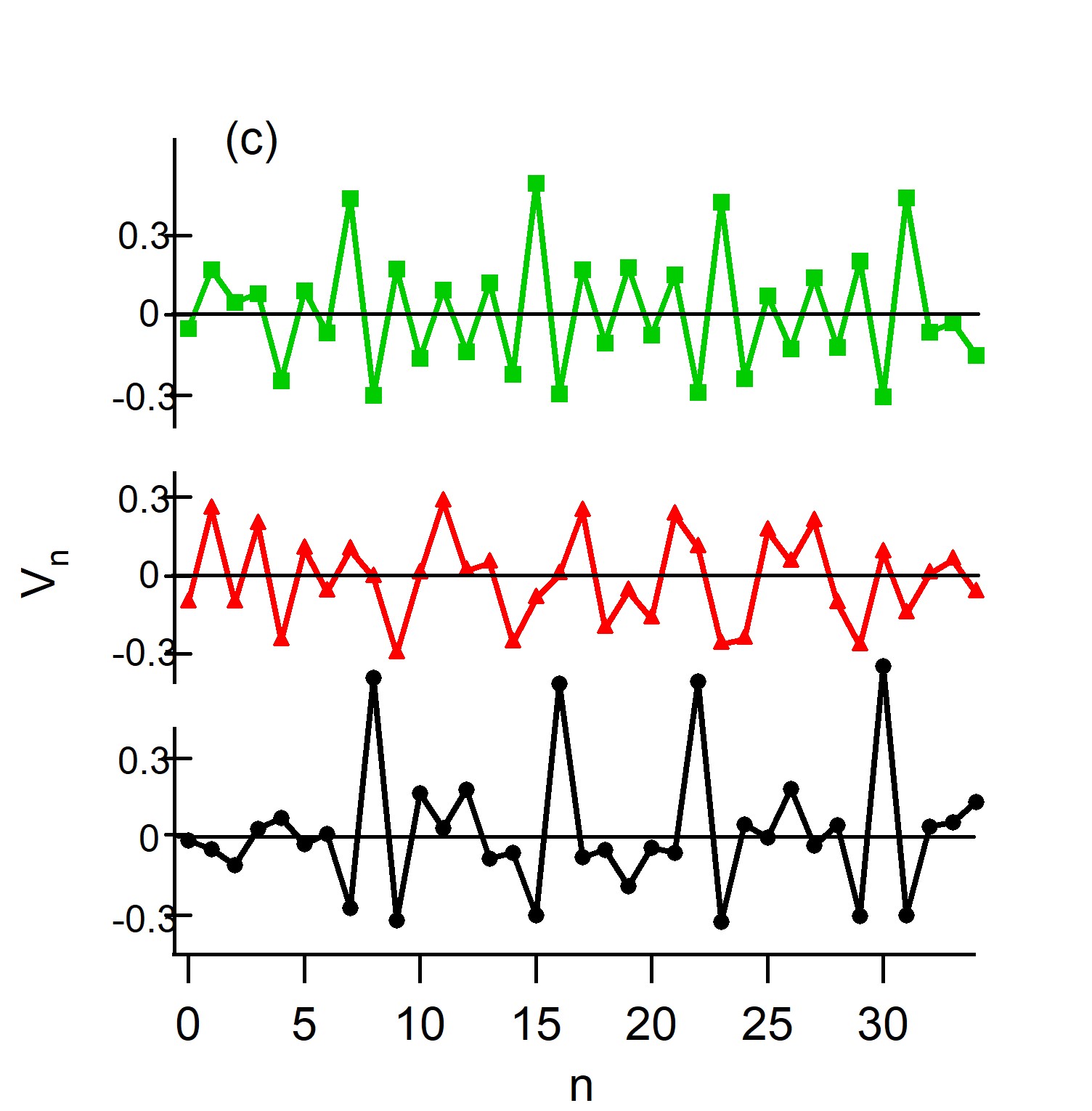}
\caption{(a) Experimental data for $f_d$=500 kHz and amplitude 1.6 V. We see multiple ZB ILMs. (b) Numerical simulation with the same driving voltage, and a frequency of $\omega=539$ kHz. (c) Numerical voltage profile at three different times, $\tau$=3010, 3011, 3012.} \label{num2}
\end{figure}

\begin{figure}[tbh]
\centering
\includegraphics[width=3.0in]{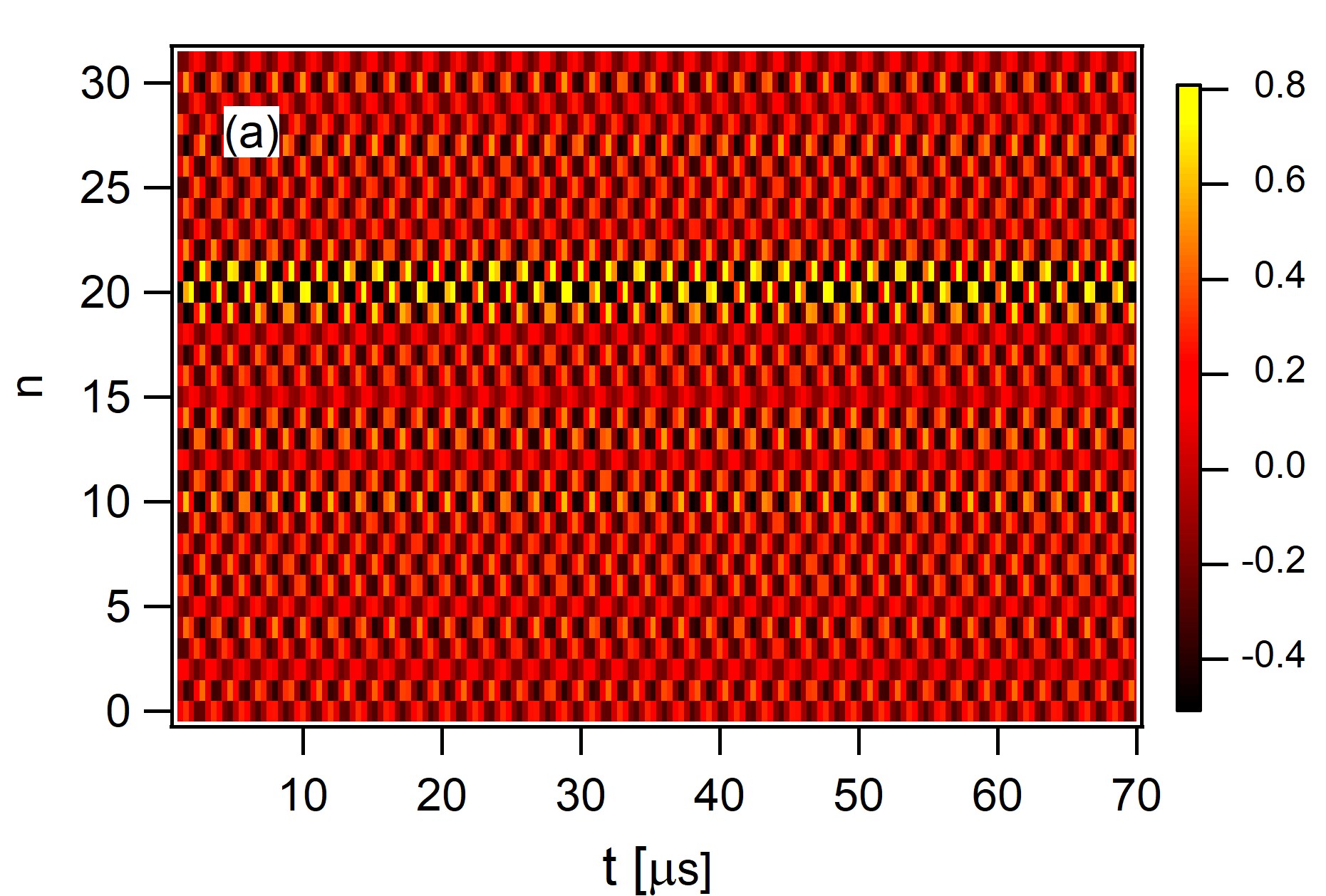}
\includegraphics[width=2.6in]{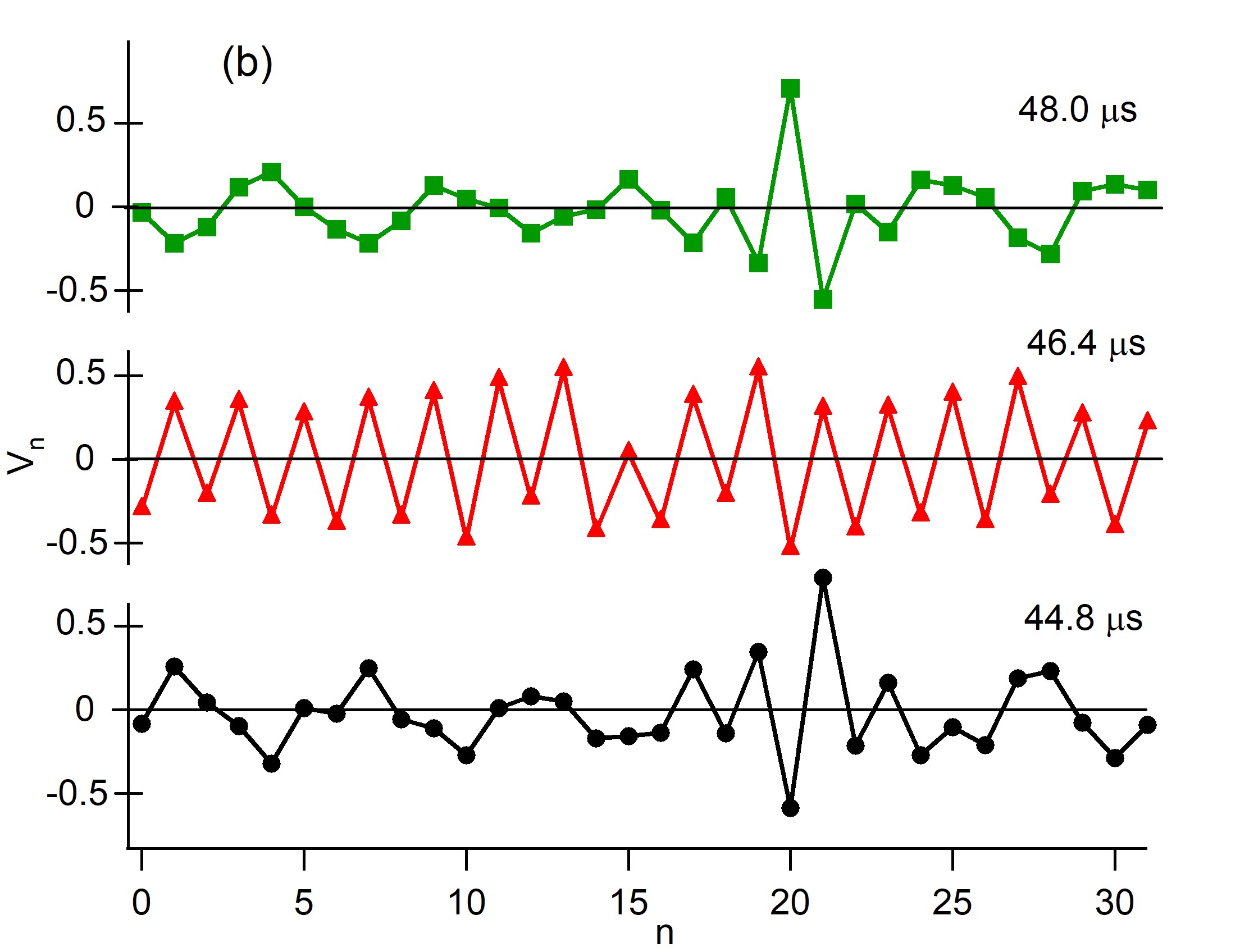}
\caption{(a) Experimental frequency is lowered to 479 kHz and amplitude raised to 2.0 V. We observe one ZB ILM persisting indefinitely. (b) Voltage profile at three times, t = 44.8, 46.4, and 48.0 $\mu$s. The resonance signature in the wings of the ILM is clearly evident in the bottom and top traces, but the ZB signature of the driver dominates in the middle trace.} \label{exp2}
\end{figure}

Next, let us turn to the zone boundary, where we switch to spatially staggered driving ($k=\pi/a$). The linear ZB mode at $k=\pi/a$ is calculated to reside at around 585 kHz. Figure \ref{num2}(a) shows the experimental results for a sinusoidal driver frequency set to 500 kHz, or $\omega_d = 0.85 \omega_{ZB}$, and an amplitude of 1.6 V. Four ZB ILMs are observed in the lattice for these driving conditions. 

Figure \ref{num2}(b) shows the corresponding numerical simulation of the excited ZB ILMs. The driver amplitude is identical, although the frequency has to be raised somewhat, here to 539 kHz. The initial displacements start from zero but with a small noise perturbation. Four stable ZB ILMs are generated via modulational instability (MI). The MI is not shown in this figure. In Fig.~\ref{num2}(c), the ILMs centered at node 7,8, as well as 30,31, in the simulation are clearly identifiable as having odd symmetry in the lowest panel. In the middle panel, after some elapsed time within the same period, the staggered driver signature dominates. Then, in the top panel, the two ILMs centered about 8 and 30 have clearly changed towards an even anti-symmetric configuration. We point out that this periodic exchange between even and odd ILM symmetry continues thereafter, in accordance also with the experimental results of Fig.~\ref{num2}(a). 

In the experiment it is possible to lower the driver frequency further while increasing the amplitude in order to reduce the number of ZB ILMs. This is illustrated in Fig.~\ref{exp2}(a), where only one ILM remains. From the experimental profiles displayed in Fig.~\ref{exp2}(b), it is clear that this represents a ZB ILM with neighboring nodes out-of-phase with one another. Furthermore, the data displays clear evidence for ILM resonance with the degenerate part of the plane-wave spectrum. This is again most easily discerned in Fig.\ref{exp2}(b), which shows the voltage profile at three particular instants of time. We see the ILM
with a symmetric (top panel) and anti-symmetric (bottom panel) configuration
of the voltages, while the middle panel transitions between the two.
Note also that the plane-wave signature is evident within the wings of the ILM in both the top and bottom panel, yet the nature of the configuration
is less clear
in the middle panel. 

\begin{figure}[tbh]
\centering
\includegraphics[width=3.0in]{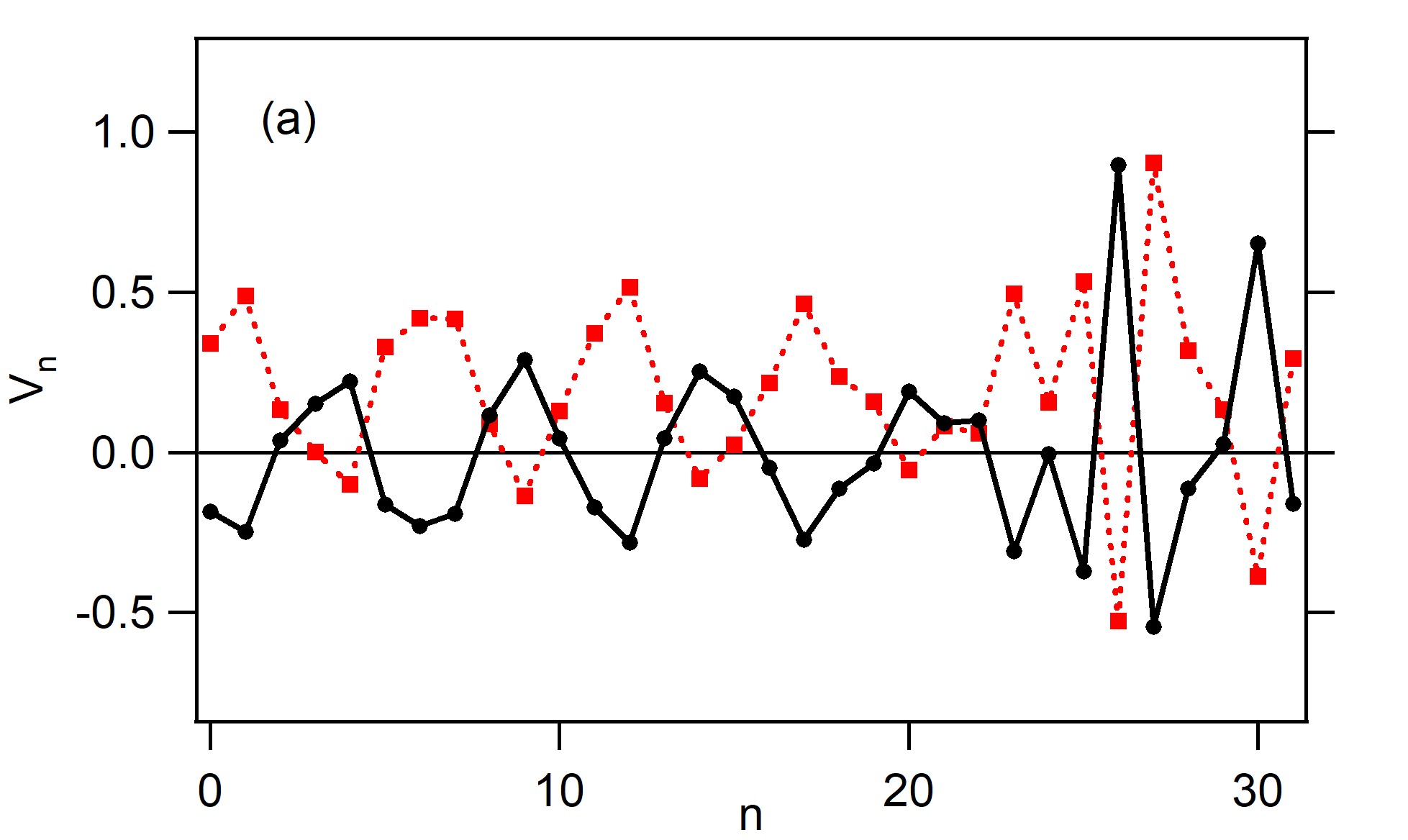}
\includegraphics[width=3.0in]{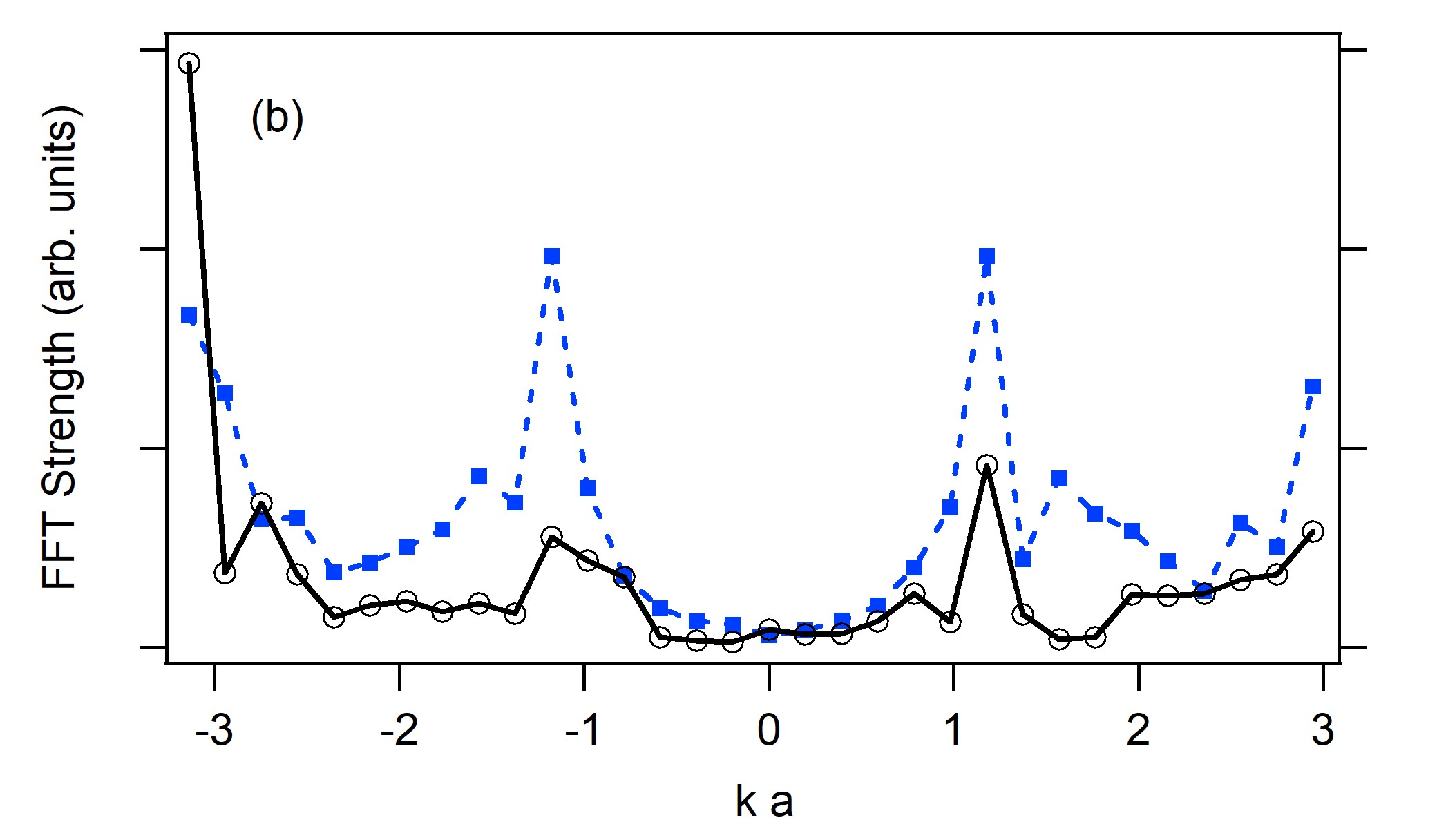}
\caption{(a) Experimental node voltages for subharmonic driving at a frequency of 980 kHz with an amplitude of 9.5 V. We observe one main resonant, subharmonically driven ILM. The voltage profiles at two instants of time, 42.8 (solid) and 43.6 (dotted) $\mu$s.(b) The spatial Fourier amplitudes computed from the experimental data. The dashed trace with square markers (blue) is computed from the data set in (a). Circles (black) derive from data in Fig.~\ref{exp2}. We see that subharmonic generation of resonant ILMs enhances the oscillatory background.} 
\label{exp3}
\end{figure}

A spatially staggered driving signal can ordinarily lock onto any ZB mode, and here we find that it can also stabilize a resonant ILM. A question is whether such ILMs can also be excited via spatially uniform, but subharmonic driving. It is known that subharmonic driving, being itself a nonlinear process, can project out into modes within the linear band, thus potentially coupling to a ZB ILM. 
Indeed, we find that the resonant ILM at the ZB can be excited in this manner. The results are shown in Fig.~\ref{exp3}(a). Here the driver frequency was set to 980 kHz with a large amplitude of 9.5V and was of square-wave shape. One main ILM centered at nodes 26 and 27 is generated (along with some smaller energy localization at n = 30). However, we now also see an enhanced plane-wave signature from the degenerate part of the dispersion curve.

To analyze the difference between the two types of driving, we compute from the experimental data, $V(n,t)$, the two-dimensional FFT, which yields a Fourier amplitude as a function of frequency and wavenumber. We then select the slice corresponding to the fixed oscillation frequency of the ILM. The result is displayed in Fig.~\ref{exp3}(b), where the dashed, blue line (square markers) corresponds to the subharmonic driver, and the solid, black line (circles) to a direct driver. The latter is computed from the data in Fig.~\ref{exp2}.
It is clear that the peak response heights are distinct for the two types of driving. While direct driving results in the greatest amplitude at the ZB ($k=-\pi/a$), subharmonic driving deposits slightly more energy into the degenerate plane waves than the ZB ILM. This may not be surprising since a subharmonic driver can presumably excite these modes directly and not just indirectly via the ILM resonance with these plane waves.  

\section{Conclusions \& Future Work}
We have demonstrated experimentally that two types of long-lived
(enough to also be experimentally observed)
ILMs exist and can be generated in an electrical lattice with the addition
of second-neighbor couplings: standard and resonant ILMs, also known as
nanopterons. The former appear at the ZC, whereas the latter at the ZB. We
have furthermore investigated these ILMs numerically using a simplified model
of the varactor diodes and identified suitable
initial guesses for such breathers, based on the undamped-undriven
simplified model within the rotating wave approximation.  

In the experiment, resonant ILMs can be preserved over the course
of the dynamical evolution by either direct or subharmonic driving. In the simplified model without gain and loss, the resonant ILM may scatter its
energy into the degenerate plane-waves and be thus led to decay. However,
more realistic numerical simulations that incorporate damping and driving
show that resonant ILMs can be preserved  at the ZB due to the action
of the periodic driving. We conclude that both at the zone boundary as well as the zone center, ILMs can be found and stabilized by the driver, in good qualitative agreement with the the RWA-predicted structures and the
corresponding numerical simulations.

Nevertheless, numerous open questions have also emerged in this study.
For the ZC modes, it does not elude us that their harmonics find themselves
inside the band of linear excitations. Without dissipation, this would clearly represent a pathway detrimental for the longevity of these localized structures. 
The same is true for the ZB ILMs which in the undamped/undriven model
can, in principle, be resonantly decaying towards extended state
excitations. On the other hand, as indicated above, in the driven/damped
variant of the system, the breathers appear to be robust. Nevertheless,
a natural question is whether they can constitute exact time-periodic
solutions of the system or not, and perhaps even more importantly 
(from the relevant Floquet multiplier analysis of the corresponding
monodromy matrix) whether they will be genuinely stable or not.
A continuation over the frequencies may reveal possible variations and
bifurcations along the relevant branch, potentially revealing the
source of the observed mobility of the ZC structures, as the frequency
is decreased. Numerous additional topics can be considered in further
detail, including also the modulational stability in a semi-exact analytical
form: although this may not be possible to perform in the full breather
setting, it should definitely be tractable in a monochromatic discrete
nonlinear Schr{\"o}dinger-like approximation of the model.
A more systematic classification of the unstable modes and the corresponding
byproduct of the MI may well prove helpful towards identifying the
number of instability-induced breathers for each frequency. All these
are important questions in this emerging field of the study of highly
nonlinearly coupled chains involving both first and second-neighbors.
As such they will be considered in future studies on the subject.

{\it Acknowledgements}. PGK acknowledges
that this work was made possible
by NPRP grant \# [9-329-1-067] from Qatar National Research Fund (a member of Qatar Foundation). The findings
achieved herein are solely the responsibility of the authors.

\end{document}